\documentclass[]{jpsj2}

\title{Faster Annealing Schedules for Quantum Annealing}
\author{Satoshi Morita}

\inst{Department of Physics, Tokyo Institute of Technology, Oh-okayama,
Meguro-ku, Tokyo 152-8551}

\abst{New annealing schedules for quantum annealing are proposed based
on the adiabatic theorem. These schedules exhibit faster decrease of the
excitation probability than a linear schedule. To derive this
conclusion, the asymptotic form of the excitation probability for
quantum annealing is explicitly obtained in the limit of long annealing
time. Its first-order term, which is inversely proportional to the
square of the annealing time, is shown to be determined only by the
information at the initial and final times.  Our annealing schedules
make it possible to drop this term, thus leading to a higher order
(smaller) excitation probability.  We verify these results by solving
numerically the time-dependent Schr\"{o}dinger equation for small size
systems}

\kword{quantum annealing, annealing schedule, adiabatic theorem,
quantum algorithm, optimization problem}

\begin{document}
\maketitle
\section{Introduction}

Quantum annealing (QA) is a novel generic algorithm to solve
optimization problems. \cite{KN,DC,ST} As can be seen from the fact that
QA is also called quantum adiabatic evolution in the field of quantum
computation, \cite{FGGS} QA is based on the adiabatic theorem of quantum
mechanics. \cite{Mess} One considers the time-dependent Hamiltonian
consisting of two terms, the potential term and kinetic energy term, the
latter being introduced by hand to solve the minimization problem of the
former. The Hamiltonian initially has only the kinetic energy term and
smoothly changes to eventually consist only of the potential term at the
end of a process.  If the Hamiltonian varies slowly, the adiabatic
theorem guarantees that the system follows the ground state of the
instantaneous Hamiltonian during the process.  The idea of QA is that
the non-trivial target, the ground state of the potential energy, is
obtained after the adiabatic evolution starting from the trivial ground
state of the kinetic energy.

An important question is how the error rate of QA depends on the
annealing time $\tau$. If $\tau$ is too short, the change of the
Hamiltonian is rapid and then the non-adiabatic transition may
occur. Consequently the algorithm would miss the desired solution.  An
earlier discussion suggested that the residual energy defined by the
energy difference between the solution obtained by an algorithm and true
one is proportional to $(\log \tau)^{-\zeta}$ with $\zeta\sim
6$. \cite{SMTC} While this result is still under debate, \cite{SPA,SO}
the asymptotic behavior of the excitation probability in the large $\tau$
limit is well-established. \cite{SO} According to the adiabatic theorem,
\cite{Mess} the excitation probability after the adiabatic evolution obeys
the power law, $\tau^{-2}$.

To implement QA and derive the $\tau^{-2}$ error rate, one ordinarily
assumes that the Hamiltonian depends linearly on time. However, there is
no theoretical restriction on the annealing schedule, the changing rate
of the Hamiltonian. In the present paper, we derive the explicit
expression of the coefficient of the $\tau^{-2}$ term in the excitation
probability without this restriction of linearity. It is surprising that
its upper bound is determined only by the information at the initial and
final times.  Based on this result, we find the condition that this
coefficient vanishes and propose new faster annealing schedules having a
much faster rate $\tau^{-2m}$ of error decrease.

This article is organized as follows. In the next section, the upper
bound of the excitation probability is derived for a general quantum
system. We apply in \S 3 this result to QA and propose faster
annealing schedules. In \S 4, we verify them by numerical simulations
for the two-level system, the transverse field Ising model and the
database search problem. The last section is devoted to conclusion.

\section{Upper bound for excitation probability}
Let us consider the general Hamiltonian which depends on time $t$ only
through the dimensionless time $s=t/\tau$,
\begin{equation}
 H(t)\equiv\tilde{H}(t/\tau)=\tilde{H}(s),
\end{equation}
where $\tau$ is the total annealing time.  The state vector follows the
Schr\"{o}dinger equation,
\begin{equation}
 {\rm i}\frac{\rm d}{{\rm d}t}|\psi(t)\rangle = H(t)|\psi(t)\rangle,
  \label{eq:Schrodinger}
\end{equation}
or, in terms of the dimensionless time,
\begin{equation}
 {\rm i}\frac{\rm d}{{\rm d}s}|\tilde{\psi}(s)\rangle 
  = \tau \tilde{H}(s)|\tilde{\psi}(s)\rangle,
\end{equation}
where we set $\hbar=1$. We assume that the initial state is chosen to be
the ground state of $\tilde{H}(0)$ and that the ground state of
$\tilde{H}(s)$ is not degenerate for $0\leq s\leq 1$. Thus the state
vector is expected to keep track of the instantaneous ground state of
$\tilde{H}(s)$ for sufficiently large $\tau$. The adiabatic theorem
\cite{Mess} provides a condition for the adiabatic evolution as
\begin{equation}
 \tau \gg \max_{s,j} \left\{A_j(s)\right\},
  \label{eq:AC}
\end{equation}
\begin{equation}
 A_j(s)\equiv \frac{1} {\Delta_j(s)^2}
  \left| \left\langle j(s) \right|\frac{{\rm d}\tilde{H}(s)}{{\rm d}s}
   \left|0(s) \right\rangle \right|,
\end{equation}
where $|j(s)\rangle$ denotes the $j$th instantaneous eigenstate of
$\tilde{H}(s)$ with the eigenvalue $\varepsilon_j(s)$ and
$\Delta_j(s)\equiv \varepsilon_j(s)-\varepsilon_0(s)$. We assume that
$|0(s)\rangle$ is the ground state of $\tilde{H}(s)$.

Under the above adiabatic condition (\ref{eq:AC}), the asymptotic
expansion with respect to $\tau$ yields the excitation amplitude at
$s=1$ as
\begin{equation}
 \langle j(1)|\tilde{\psi}(1)\rangle\simeq \int_{0}^{1}{\rm d}s
 \frac{{\rm e}^{{\rm i}\tau\int_{0}^{s}{\rm d}s'\,\Delta_j(s')}}
 {\Delta_j(s)}\left\langle j(s)\right|
 \frac{{\rm d}\tilde{H}(s)}{{\rm d}s}\left|0(s)\right\rangle 
 +O(\tau^{-2}).
\end{equation}
Using the integration by parts, we obtain the upper bound for the excitation
probability as
\begin{equation}
 \left|\langle j(1)|\tilde{\psi}(1)\rangle\right|^2\lesssim
  \frac{1}{\tau^2}\left\{A_j(0)+A_j(1)\right\}^2+O(\tau^{-3}).
\end{equation}
This formula implies that the coefficient of the $\tau^{-2}$ term is
determined only by the state of the system at $s=0$ and 1 and vanishes if
$A_j(s)$ is zero at $s=0$ and 1.

When the $\tau^{-2}$ term vanishes, a similar calculation yields the
next order term of the excitation probability. If
$\tilde{H}'(0)=\tilde{H}'(1)=0$, we obtain
\begin{equation}
 \left|\langle j(1)|\tilde{\psi}(1)\rangle\right|^2\lesssim
  \frac{1}{\tau^4}\left\{A_j^{(2)}(0)+A_j^{(2)}(1)\right\}^2
  +O(\tau^{-5}),
\end{equation}
where we defined
\begin{equation}
 A_j^{(m)}(s)\equiv \frac{1}{\Delta_j(s)^{m+1}} \left|\langle j(s)|
 \frac{{\rm d}^m\tilde{H}(s)}{{\rm d}s^m}|0(s)\rangle\right|.
\end{equation}
It is easy to see that the $\tau^{-4}$ term also vanishes when
$\tilde{H}''(0)=\tilde{H}''(1)=0$. We generalize these results as
follows: If the $k$th derivative of $\tilde{H}(s)$ is equal to zero at
$s=0$ and 1 for all $k=1,2,\cdots, m-1$, the excitation probability has
the upper bound
\begin{equation}
 \left|\langle j(1)|\tilde{\psi}(1)\rangle\right|^2\lesssim
  \frac{1}{\tau^{2m}}\left\{A_j^{(m)}(0)+A_j^{(m)}(1)\right\}^2
  +O(\tau^{-2m-1}),
  \label{eq:UB}
\end{equation}
which is proportional to $\tau^{-2m}$.

\section{Faster annealing schedules}
Although we have so far considered the general time-dependent
Hamiltonian, the ordinary Hamiltonian for QA is composed of the
potential term and the kinetic energy term,
\begin{equation}
 \tilde{H}(s)=\left\{1-f(s)\right\}H_{\rm kin}
  +f(s) H_{\rm pot}.\label{eq:Hamiltonian}
\end{equation}
The function $f(s)$, representing the annealing schedule, satisfies
$f(0)=0$ and $f(1)=1$. Thus $\tilde{H}(0)=H_{\rm kin}$ and
$\tilde{H}(1)=H_{\rm pot}$. The ground state of $H_{\rm pot}$
corresponds to the optimal solution for the optimization problem. The
kinetic energy is chosen so that its ground state is trivial. Therefore
the above Hamiltonian connects the trivial initial state and the
non-trivial desired solution.

The condition for the $\tau^{-2m}$ term to exist is obtained
straightforwardly from the results of the previous section because the
Hamiltonian (\ref{eq:Hamiltonian}) depends on time only through the
annealing schedule $f(s)$. It is sufficient that the $k$th derivative of
$f(s)$ is zero at $s=0$ and 1 for $k=1,2,\cdots, m-1$. We note that
$f(s)$ should belong to $C^m$, that is, $f(s)$ is an $m$th differentiable
function whose $m$th derivative is continuous.

Examples of the annealing schedules $f_m(s)$ with the $\tau^{-2m}$ error
rate are the following polynomials:
\begin{equation}
 f_1(s)=s,\label{eq:f1}
\end{equation}
\begin{equation}
 f_2(s)=s^2 (3-2s),
\end{equation}
\begin{equation}
 f_3(s)=s^3 (10-15s+6s^2),
\end{equation}
\begin{equation}
 f_4(s)=s^4 (35-84s+70s^2-20s^3).\label{eq:f4}
\end{equation}
The linear annealing schedule $f_1(s)$, which shows the $\tau^{-2}$
error, has ordinarily been used in the past studies. \cite{DC,ST}
Although we here list only polynomials symmetrical with respect to the
point $s=1/2$, this is not essential. For example, $f(s)=(1-\cos (\pi
s^2))/2$ also has the $\tau^{-4}$ error rate because
$f'(0)=f'(1)=f''(0)=0$ but $f''(1)= -2\pi^2$.

\section{Numerical results}

To confirm the upper bound for the excitation probability discussed in
the previous section, it is instructive to study the two-level system
with the Hamiltonian,
\begin{equation}
 H_{\rm LZ}(t)=-\left(\frac{1}{2}-f\left(\frac{t}{\tau}\right)\right) 
 h\sigma^z-\alpha \sigma^x,
 \label{eq:H_LZ}
\end{equation}
where the $\sigma^{\alpha}$ $(\alpha=x,y,z)$ are the Pauli matrices.
The energy gap of $H_{\rm LZ}$ has the minimum $2\alpha$ at
$f(s)=1/2$. If the annealing time $\tau$ is not large enough to satisfy
eq. (\ref{eq:AC}), the non-adiabatic transition occurs. The Landau-Zener
theorem \cite{Landau, Zener} provides the excitation probability $P_{\rm
ex}=\left|\langle 1(1)|\tilde{\psi}(1)\rangle\right|^2$ as
\begin{equation}
 P_{\rm ex} =\exp\left(-\frac{\pi \alpha^2 \tau}{f'(s^*) h}\right),
  \label{eq:p1_LZ}
\end{equation}
where $s^*$ denotes a solution for $f(s^*)=1/2$.  On the other hand, if
$\tau$ is sufficiently large, the system evolves adiabatically during
the process. Thus the excitation probability has the upper bound
(\ref{eq:UB}), which is estimated as
\begin{equation}
 P_{\rm ex} \lesssim \frac{4h^2\alpha^2}{\tau^{2m}(h^2+4\alpha^2)^{m+2}}
  \left\{\left|\frac{{\rm d}^m f}{{\rm d}s^m}(0)\right|
   +\left|\frac{{\rm d}^m f}{{\rm d}s^m}(1)\right|\right\}^2.
  \label{eq:p2_LZ}
\end{equation}

\begin{figure}
 \begin{center}
  \includegraphics{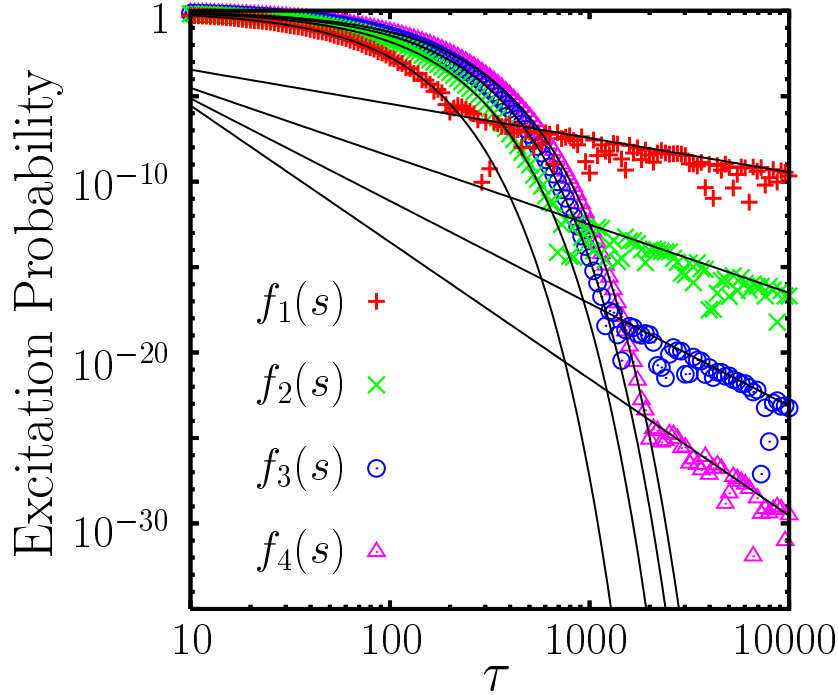} 
  \caption{The annealing-time dependence of the
  excitation probability for the two-level system (\ref{eq:H_LZ}) using
  schedules (\ref{eq:f1}) to (\ref{eq:f4}). The
  curved and straight lines show eqs. (\ref{eq:p1_LZ}) and
  (\ref{eq:p2_LZ}) for each annealing schedule, respectively. The
  parameters in eq. (\ref{eq:H_LZ}) are chosen to be $h=2$ and
  $\alpha=0.2$.}
  \label{fig:LZ}
 \end{center} 
\end{figure}

We numerically solved the Schr\"{o}dinger equation
(\ref{eq:Schrodinger}) for this system (\ref{eq:H_LZ}) with the
Runge-Kutta method. Figure \ref{fig:LZ} shows the result for the
excitation probability with annealing schedules
(\ref{eq:f1})-(\ref{eq:f4}). The initial state is the ground state of
$H_{\rm LZ}(0)$. The parameters are chosen to be $h=2$ and $\alpha=0.2$. The
curved and straight lines show eqs. (\ref{eq:p1_LZ}) and
(\ref{eq:p2_LZ}), respectively.  In the small and large $\tau$ regions, the
excitation probability perfectly fits those two expressions.

We carried out simulations of a rather large system, the Ising spin
system with random interactions. The quantum fluctuations are introduced
by the uniform transverse field. Thus, the potential and kinetic energy
terms are defined by
\begin{equation}
 H_{\rm pot}=-\sum_{\langle ij \rangle}J_{ij}\sigma_i^z\sigma_j^z
  -h\sum_{i=1}^{N} \sigma_i^z,
\end{equation}
\begin{equation}
 H_{\rm kin}=-\Gamma\sum_{i=1}^{N}\sigma_i^x.
\end{equation}
The initial state, namely the ground state of $H_{\rm kin}$, is the
all-up state along the $x$ axis,
\begin{equation}
 |\psi(0)\rangle = \bigotimes_{i} \frac{|z+\rangle_i+|z-\rangle_i}{\sqrt{2}},
\end{equation}
where $|z+\rangle_i$ and $|z-\rangle_i$ stand for the eigenstates of
$\sigma_i^z$ with eigenvalues $1$ and $-1$, respectively.

The residual energy $E_{\rm res}$, the energy difference between the
solution obtained by the QA process and the exact one, is a useful
measure for the error rate of QA. It has the same behavior as the
excitation probability because it is rewritten as
\begin{align}
 E_{\rm res} &\equiv\langle \tilde{\psi}(1)| H_{\rm pot}
  |\tilde{\psi}(1)\rangle-\varepsilon_0(1)\\
 &=\sum_{j> 0} \Delta_j(1)
 \left|\langle j(1) | \tilde{\psi}(1)\rangle \right|^2.
\end{align}
Therefore $E_{\rm res}$ is expected to be asymptotically in proportion
to $\tau^{-2m}$ using faster annealing schedules.

\begin{figure}
 \begin{center}
  \includegraphics{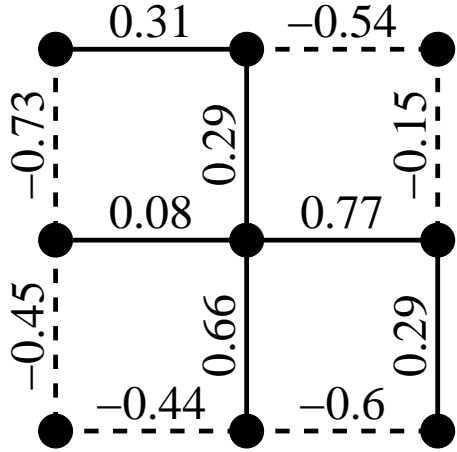}
  \caption{Configuration of random interactions $\{J_{ij}\}$ on the
  $3\times 3$ square lattice which we investigated. The solid and
  dashed lines indicate ferromagnetic and antiferromagnetic
  interactions, respectively. }
  \label{fig:sg_int}
 \end{center} 
\end{figure}

\begin{figure}
 \begin{center}
  \includegraphics{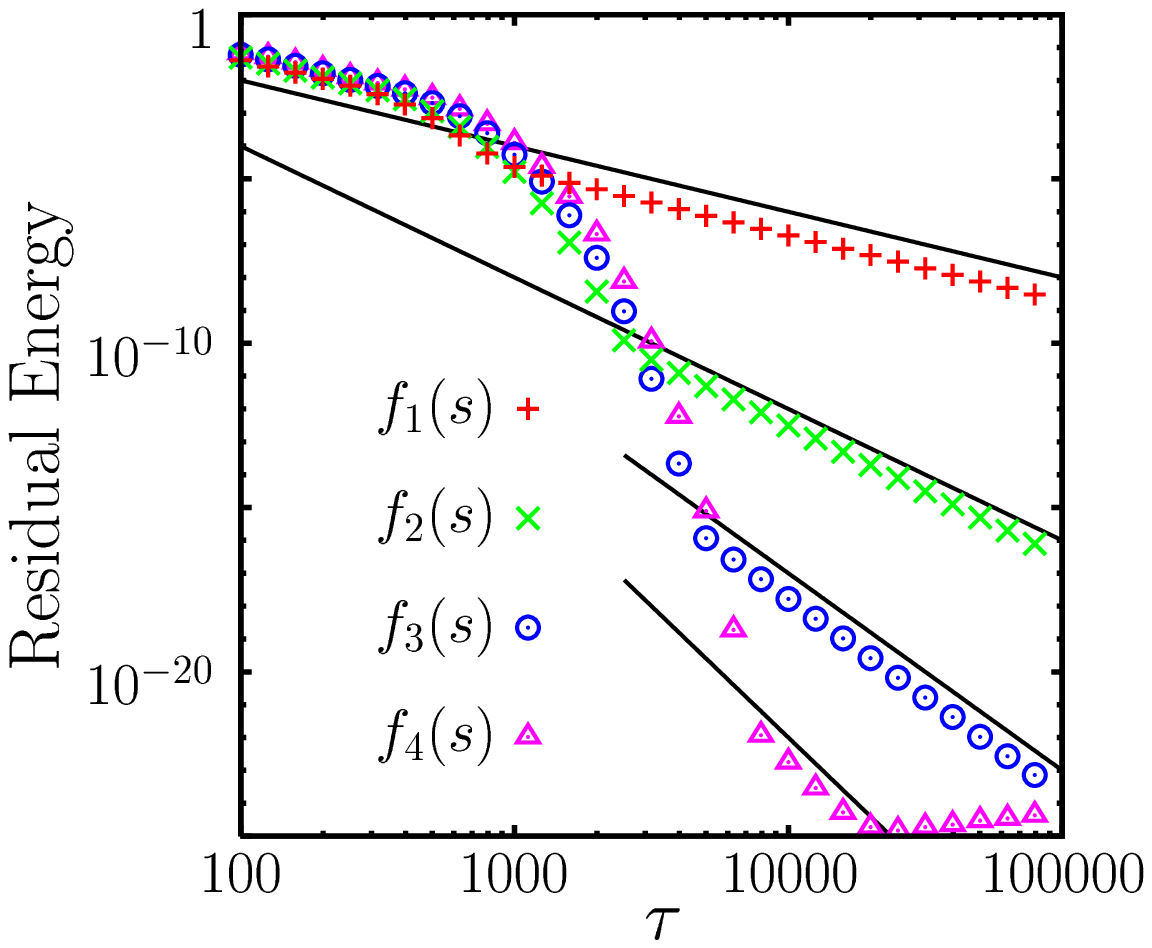}
  \caption{The annealing-time dependence of the residual energy for the
  two-dimensional spin glass model with faster annealing schedules. The
  solid lines stand for functions proportional to $\tau^{-2m}$
  $(m=1,2,3,4)$. The parameters are $h=0.1$ and $\Gamma=1$.}
  \label{fig:sg}
 \end{center} 
\end{figure}

We investigated the two-dimensional square lattice of size $3\times
3$. The quenched random coupling constants $\{J_{ij}\}$ are chosen from
the uniform distribution between $-1$ and $+1$, as shown in
Fig. \ref{fig:sg_int}. The parameters are $h=0.1$ and $\Gamma=1$.
Figure \ref{fig:sg} shows the $\tau$ dependence of the residual energy
using annealing schedules (\ref{eq:f1})-(\ref{eq:f4}). Straight lines
representing $\tau^{-2m}$ $(m=1,2,3,4)$ are also shown for comparison.
The data clearly indicates the $\tau^{-2m}$-law for large $\tau$. Note
that the irregular behavior around $E_{\rm res}\sim 10^{-25}$ comes from
numerical rounding errors.

Next, we apply the faster annealing schedule to the database search
problem, finding an item in an unsorted database. Consider $N$ items,
among which one is marked. The goal of this problem is to find the
marked item in a minimum time. The pioneering quantum algorithm proposed
by Grover \cite{Grov} solves this task in time of order $\sqrt{N}$,
whereas the classical algorithm tests $N/2$ items on average. Farhi {\it
et al.} \cite{FGGS} proposed a QA version of Grover's algorithm and
Roland and Cerf \cite{RC} found an annealing schedule which has the same
complexity as Grover's algorithm. Although their schedule is
optimal in the sense that the excitation probability by the adiabatic
transition is equal to a small constant at each time, it has the
$\tau^{-2}$ error rate. We show that annealing schedules with the
$\tau^{-2m}$ error rate can be constructed by a slight modification of
their optimal schedule.

Let us consider the Hilbert space which has the basis states
$|i\rangle$ $(i=1,2,\cdots, N)$, and the marked state is denoted by
$|m\rangle$. Suppose that we can construct the Hamiltonian
(\ref{eq:Hamiltonian}) with two energy terms,
\begin{equation}
 H_{\rm pot}=1-|m\rangle \langle m|,
\end{equation}
\begin{equation}
 H_{\rm kin}=1-\frac{1}{N}\sum_{i=1}^{N}\sum_{j=1}^{N}|i\rangle\langle j|.
\end{equation}
The Hamiltonian $H_{\rm pot}$ can be applied without the explicit knowledge
of $|m\rangle$, which is the same assumption as in Grover's algorithm.
The initial state is a superposition of all basis states,
\begin{equation}
 |\psi(0)\rangle = \frac{1}{\sqrt{N}}\sum_{i=1}^{N}|i\rangle,
\end{equation}
which does not depend on the marked state. The energy gap between the
ground state and the first excited state,
\begin{equation}
 \Delta_1(s)=\sqrt{1-4\frac{N-1}{N}f(s)(1-f(s))},
\end{equation}
has a minimum at $f(s)=1/2$. The highest eigenvalue
$\varepsilon_2(s)=1$ is $(N-2)$-fold degenerate.

\begin{figure}
 \begin{center}
  \includegraphics{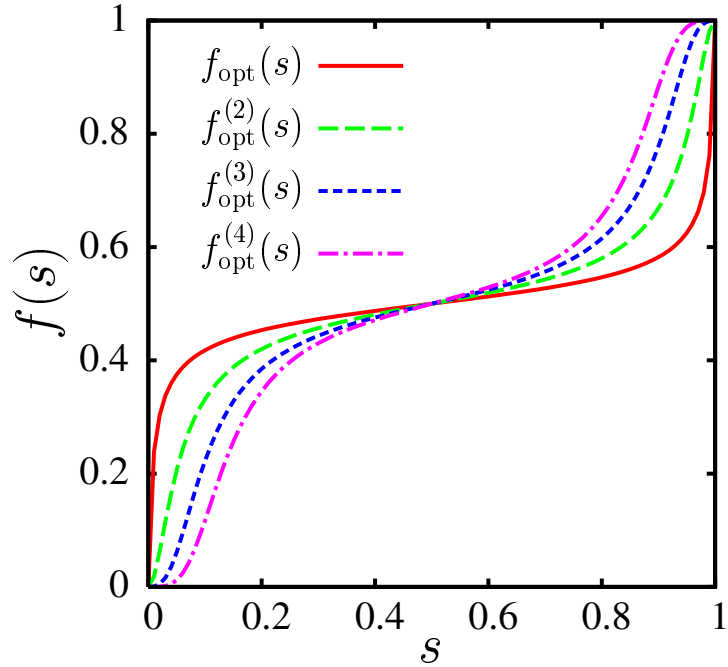}
  \caption{The optimal annealing schedules for
  the database search problem ($N=64$). The solid line denotes the
  original optimal schedule (\ref{eq:opt}) and the dashed lines stand
  for modified schedules.}  \label{fig:opt}
 \end{center} 
\end{figure}

To derive the optimal annealing schedule, we briefly review the results
reported by Roland and Cerf.\cite{RC} When the energy gap is small, a
non-adiabatic transition is likely to occur. Thus we need to change the
Hamiltonian carefully. On the other hand, when the energy gap is not very
small, too slow a change wastes time.  Therefore the annealing schedule
should be tuned to satisfy the adiabatic condition (\ref{eq:AC}) in each
infinitesimal time interval, that is,
\begin{equation}
 \frac{A_1(s)}{\tau}=\delta,
\end{equation}
where $\delta$ is a small constant.  In the database search problem,
this condition is rewritten as
\begin{equation}
 \frac{\sqrt{N-1}}{\tau N \Delta_1(s)^3}\frac{{\rm d}f}{{\rm d}s}=\delta.\label{eq:db_adia}
\end{equation}
After integration under boundary conditions $f(0)=0$ and $f(1)=1$, we
obtain that
\begin{equation}
 f_{\rm opt}(s)=\frac{1}{2}+\frac{2s-1}{2\sqrt{N-(N-1)(2s-1)^2}}.
  \label{eq:opt}
\end{equation}
As plotted by a solid line in Fig. \ref{fig:opt}, this function changes
most slowly when the energy gap takes a minimum value.  It is noted that the
annealing time is determined by the small constant $\delta$ as
\begin{equation}
 \tau=\frac{\sqrt{N-1}}{\delta},
\end{equation}
which means that the computation time is of order $\sqrt{N}$ similarly
to Grover's algorithm.

\begin{figure}
 \begin{center}
  \includegraphics{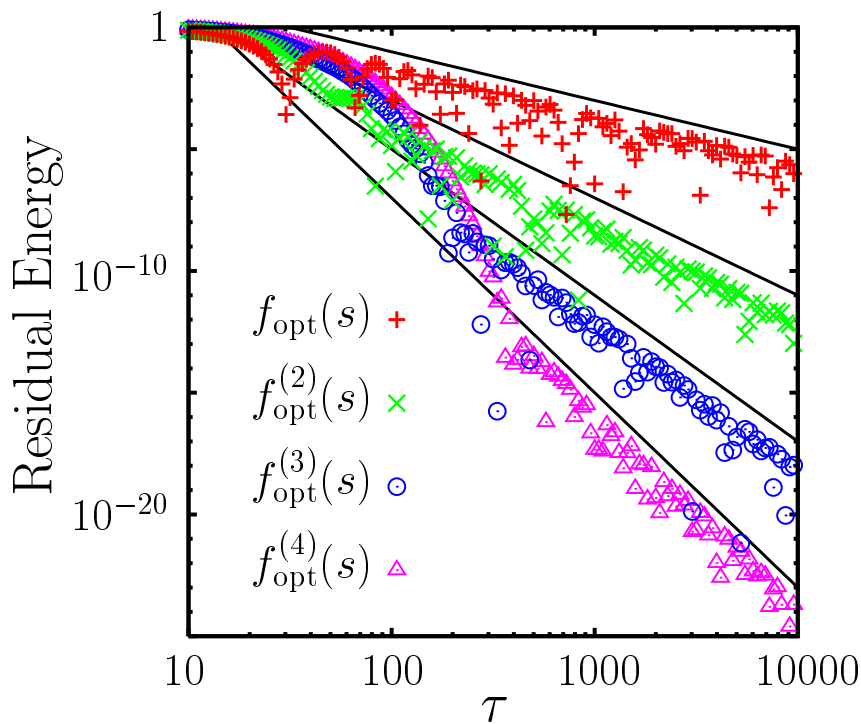}
  \caption{The annealing-time dependence of the residual
  energy for the database search problem ($N=64$) with the optimal
  annealing schedules plotted in Fig. \ref{fig:opt}. The solid lines
  stand for functions proportional to $\tau^{-2m}$ $(m=1,2,3,4)$.}
  \label{fig:db}
 \end{center} 
\end{figure}


The optimal annealing schedule (\ref{eq:opt}) shows the $\tau^{-2}$
error rate because its derivative is not zero at $s=0$ and 1.  It is
easy to see from eq. (\ref{eq:opt}) that a simple replacement of $s$
with $f_m(s)$ fulfills the condition for the $\tau^{-2m}$ error
rate. We carried out numerical simulations for $N=64$ with such
annealing schedules, $f_{\rm opt}^{(m)}(s)\equiv f_{\rm
opt}\left(f_m(s)\right)$, as plotted by dashed lines in
Fig. \ref{fig:opt}. As shown in Fig. \ref{fig:db}, the residual energy
with $f_{\rm opt}^{(m)}(s)$ is proportional to $\tau^{-2m}$.  The
characteristic time $\tau_c$ for the $\tau^{-2m}$ error rate to show up
increases with $m$: Since the modified optimal schedule $f_{\rm
opt}^{(m)}(s)$ has a steeper slope at $s=1/2$ than $f_{\rm opt}(s)$, a
longer annealing time is necessary to satisfy the adiabatic condition
(\ref{eq:db_adia}). Nevertheless, the difference in slopes of $f_{\rm
opt}^{(m)}(s)$ is only a factor of $O(1)$, and therefore $\tau_c$ is
still scaled as $\sqrt{N}$.

\section{Conclusion}

In this paper, we have studied the error rate of QA based on the
adiabatic theorem and proposed annealing schedules which present a
faster decrease of the error rate with the annealing time. The point is
that the excitation probability is determined only by the information at
$t=0$ and $\tau$ when the system evolves adiabatically from the
beginning to the end. Thus one can easily remove the usual $\tau^{-2}$
error by tuning the annealing schedule. As examples of such schedules,
we listed a few polynomials. 

Although our new faster schedules do not improve the characteristic time
$\tau_c$ for the adiabatic evolution, these schedules can be used to
produce a new faster schedule from the known efficient schedule with
small $\tau_c$. In general, it is difficult to obtain the optimal
annealing schedule analytically. Nevertheless, if one once finds an
efficient schedule for a specific problem, schedules with smaller error
rates are immediately constructed by a slight modification under the
present scheme.


\section*{Acknowledgements}
I would like to thank Profs. H.~Nishimori and G.~E.~Santoro for useful
comments and discussions. I am supported by JSPS Research Fellowships
for Young Scientists.

\end{document}